# Query on Negative Temperature, Internal Interactions and Decrease of Entropy

Yi-Fang Chang

Department of Physics, Yunnan University, Kunming, 650091, China

(e-mail: yifangchang1030@hotmail.com)

Abstract: After negative temperature is restated, we find that it will derive necessarily decrease of entropy. Negative temperature is based on the Kelvin scale and the condition dU>0 and dS<0. Conversely, there is also negative temperature for dU<0 and dS>0. But, negative temperature is contradiction with usual meaning of temperature and with some basic concepts of physics and mathematics. It is a question in nonequilibrium thermodynamics. We proposed a possibility of decrease of entropy due to fluctuation magnified and internal interactions in some isolated systems. From this we discuss some possible examples and theories.

## 1.Restatement of Negative Temperature

In thermodynamics negative temperature is a well-known idea proposed and expounded by Ramsey [1] and Landau [2], et al.

Ramsey discussed the thermodynamics and statistical mechanics of negative absolute temperatures in a detailed and fundamental manner [1]. He proved that "if the entropy of a thermodynamic system is not a monotonically increasing function of its internal energy, it possesses a negative temperature whenever $(\partial S / \partial U)_X$ is negative. Negative temperatures are hotter than positive temperature." He pointed out: from a thermodynamic point of view the only requirement for the existence of a negative temperature is that the entropy S should not be restricted to a monotonically increasing function of the internal energy U. In the thermodynamic equation relating TdS and dU, a temperature is

$$T = (\partial S / \partial U)_X^{-1}. \tag{1}$$

An assumption, entropy S increases monotonically with U, is not necessary in the derivation of many thermodynamic theorems. Negative temperatures are hotter than infinite temperature. Negative temperature is an unfortunate and misleading. If the temperature function had been chosen as $-1/T$, then the coldest temperature would correspond to $-\infty$ for this function, infinite temperatures on the conventional scale would correspond to 0, and negative temperatures on the conventional scale would correspond to positive values of this function.

Ramsey proposed: "One of the standard formulations of the second law of thermodynamics must be altered to the following: It is impossible to construct an engine that will operate in a closed cycle and prove no effect other than (a) the extraction of heat from a positive-temperature reservoir with the performance of an equivalent amount of work or (b) the rejection of heat into a negative-temperature reservoir with the corresponding work being done on the engine. A thermodynamic system that is in internal thermodynamic equilibrium, that is otherwise essentially isolated." Negative temperatures are applied to some mutually interacting nuclear spin system. Klein justified Ramsey's criteria for systems capable of negative absolute temperatures [3]. Only premise of Ramsey's statement is Kelvin definition (1). Even he implied that entropy might decrease with U, i.e., in an original definition of temperature

$$T=dU/dS, \tag{2}$$

when dU>0 and dS<0, T<0.

Intuitively, the physical meaning of temperature is that it describes whether a body is "hot or cold" [4]. A definition of an absolute thermodynamic temperature scale is proportional to the quantity of heat. Maxwell's definition is that the temperature of a body is its thermal state considered with reference to its power of communicating heat to other bodies, which was adopted substantially unchanged by Planck and Poincare [5]. In microscopic thermodynamics temperatures are related with the states of molecular motions. Kelvin temperature scale is defined by the relationship (1).

It is a little different that Landau proved negative temperature. In Landau's book <Statistical Physics> [2] negative temperature was stated as following: Let us consider some peculiar effects related to the properties of paramagnetic dielectrics. Here the interaction of these moments brings about a new magnetic spectrum, which is superposed on the ordinary spectrum. From this the entropy is

$$S_{mag} = N \ln g - \frac{1}{2T^2} < (E_n - \overline{E}_n)^2 >, \tag{3}$$

where N is the number of atoms, g is the number of possible orientations of an individual moment relative to the lattice, $E_n$ are the energy levels of the system of interacting moments, and $\overline{E}_n$ is the average as the ordinary arithmetic mean. This shall regard the atomic magnetic moments fixed at the lattice sites and interacting with one another as a single isolated system. Further, there has the interesting result that the system of interacting moments may have either a positive or a negative temperature. "At T=0, the system is in its lower quantum state, and its entropy is zero" [2]. In fact, the temperature T=0 (absolute zero) is impossibly achieved. "As the temperature increases, the energy and entropy of the system increase monotonically. At T=∞, the energy is $\overline{E}_n$ and the entropy reaches its maximum value Nlng; these values correspond to a distribution with equal probability over all quantum states of the system, which is the limit of the Gibbs distribution as T→∞." The statement of original negative temperature is based on the two premises: entropy of the system increase monotonically, and the Gibbs distribution holds. From this some strange arguments [2] are obtained:

(a). "The temperature T=-∞ is physically identical with T=∞; the two values give the same distribution and the same values of the thermodynamic quantities for the system." According to the general definition, temperature cannot be infinite, since the quantity of heat or molecular motion all cannot be infinite. Negative temperature, even negative infinite temperature is stranger. In the same book <Statistical Physics>, Landau proved a very important result that the temperature must be positive: T>0 [2]. Moreover, T=∞=-∞ is excluded by mathematics.

(b). "A further increase in the energy of the system corresponds to an increase in the temperature from T=∞", and "the entropy decreases monotonically."

(c). "At T=0- the energy reaches its greatest value and the entropy returns to zero, the system then being in its highest quantum state." This obeys Nernst's theorem, but in which the quantity of heat is zero at T=0, while at T=0- it possesses highest quantum state! I do not know whether T=0=T=0- holds or not.

(d). "The region of negative temperature lies not below absolute zero but above infinity", i.e., "negative temperatures are higher than positive ones".

## 2. Query on "Negative Temperature" and Decrease of Entropy

In a previous paper [6], we proved that since fluctuations can be magnified due to internal interactions under a certain condition, the equal-probability does not hold. The entropy would be defined as

$$S(t) = -k \sum_r P_r(t) \ln P_r(t).$$ (4)

From this or $S = k \ln \Omega$ in an internal condensed process, possible decrease of entropy is calculated. If various internal complex mechanism and interactions cannot be neglected, a state with smaller entropy (for example, self-organized structures) will be able to appear. In these cases, the statistics and the second law of thermodynamics should be different, in particular, for nonequilibrium thermodynamics [7,8]. Because internal interactions bring about inapplicability of the statistical independence, decrease of entropy due to internal interactions in isolated system is caused possibly. This possibility is researched for attractive process, internal energy, system entropy and nonlinear interactions, etc [6].

In fact, negative temperature derives necessarily decrease of entropy. We think, "T=∞=-∞" is only a finite threshold temperature $T_c$, which corresponds to a value of entropy from increase to decrease, and this entropy is a maximum Nlng. According to the basic equation of thermodynamics, i.e., Euler equation [9],

$$S = \frac{U}{T} - \frac{YX}{T} - \sum_i \frac{\mu_i N_i}{T}.$$ (5)

If energy is invariance, corresponding temperature should be

$$T = \frac{1}{S}(U - YX - \sum_i \mu_i N_i) \leq \frac{E}{N \ln g}.$$ (6)

This value should be testable and measurable. Such temperature and energy increase continuously, and entropy decreases to a minimum, but cannot be zero. Of course, energy passes necessarily from "negative temperature system" to positive temperature system.

Next, the Gibbs distribution is

$$w_n = A e^{-E_n/T} . \tag{7}$$

This finds "the probability $w_n$ of a state of the whole system such that the body concerned is in some definite quantum state (with energy $E_n$), i.e., a microscopically defined state", and is suitable that "the system is assumed to be in equilibrium" [2]. So long as assume that the Gibbs distribution hole always, it is necessary for negative temperature. But, in the above example and laser that is another example applied negative temperature, these states are already unstable or metastable nonequilibrium states with higher energy. Bodies of negative temperature are also completely unstable and cannot exist in Nature [2]. The entropy of a body is a function only of its internal energy [2]. In states with negative temperature, the crystal be magnetized in a strong magnetic field, then the direction of the field is reversed so quickly that the spins cannot follow it [2]. This system is in a nonequilibrium state, and its internal energy and entropy are different. Laser should be an ordering process with decrease of entropy.

Generally, the Gibbs distribution for a variable number of particles is [2]

$$w_{nN} = A e^{(\Omega + \mu N - E_{nN})/T} , \tag{8}$$

where $\Omega$ is the thermodynamic potential. From this the distributions are different for the number N of particles. The number N should be different in magnetic field with reversed direction.

The above statement of "negative temperature" proves just that entropy is able to decrease with internal interactions in an isolated system. The experimental study requires that the spin system be well isolated from the lattice system [1]. This isolation is possible if the ratio of spin-lattice to spin-spin relaxation times is large [2]. This may describes the Figure 1, which is namely Fig.1 [1] and Fig.10 [2], in which the finite threshold value $T_c$ corresponds to only a maximum point dS/dE=0.

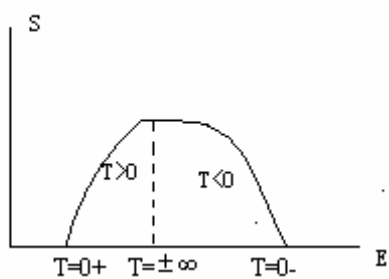

Fig.1　"negative temperature"

According to Eq.(1) or Eq.(2), when the condition dU>0 and dS<0 hold, negative temperature will be obtained. Conversely, if dU<0 and dS>0, we will also derive negative temperature. But, Eq.(2) originates from Clausius entropy dS=dQ/T, so whether above conditions could hold or not? According to Eq.(5), if $YX + \sum_i \mu_i N_i = 0$, T=U/S. In Fig.1, since U>0 (E>0) and S>0, then T>0. From this case, we cannot obtain negative temperature, and $T_c = E_c / N \ln g$.

In another book <Principles of General Thermodynamics> [5], Kelvin temperature (1) of thermodynamic system may be either positive or negative, according to whether, as the system passes through stable states with fixed parameters, the entropy increases or decreases with increasing energy. This is different with Landau's statement. In fact, this statement seems to imply that the negative temperature is unnecessary, so long as the entropy decreases with increasing energy. "A normal system can assume only positive Kelvin temperatures."±"A system at a negative Kelvin temperature is in a special state. For if this were not true we could, by definition, do work on the system adiabatically and prove that the system was at a positive Kelvin temperature. A system is capable of attaining negative Kelvin temperatures if for some of its stable states the entropy decreases for increasing energy at fixed values of the parameters."

Further, as examples, the entropy for a monatomic gas is given by

$$S = (3/2)R \ln T - R \ln \rho + S'. \tag{9}$$

Based on this, T and the density $\rho$ cannot be negative. The systems are nuclear spin ones in a pure lithium fluoride (LiF) crystal spin lattice relaxation times were as large as 5 minutes at room temperature while the spin-spin relaxation time was less than $10^{-5}$ seconds. The systems lose internal energy as they gain entropy, and the reversed deflection corresponds to induce radiation. The sudden reversal of the magnetic field produces negative temperature for the Boltzmann distribution [10]. Next, this book [5] discussed heat flow between two systems A and B at unequal temperatures, and derives

$$dQ_A (\frac{1}{T_A} - \frac{1}{T_B}) > 0. \tag{10}$$

Here let a heat quantity $dQ_A$ flow into A from B, so there should be $T_B > T_A > 0$, which is also consistent with inequality (10). Further, according to an efficiency of heat engines

$$\eta = 1 - \frac{T_2}{T_1}. \tag{11}$$

Here if either temperature is negative, the efficiency will be greater than unity. Assume that be possible, this will be testified and infinite meritorious and beneficent deeds. "The results arrived at for negative temperatures which are strange to our intuition have no practical significance in the field of energy production." But, "°s yst e m at negative Kelvin temperatures obey the second law and its many corollaries. " Of course, "it would be useless to consume work in order to produce a reservoir at a negative temperature which can be used to operate a very efficient heat engine" [5]. Therefore, this seems to imply that negative temperature is introduced only in order to obey the second law of thermodynamics.

There is the same efficiency of a Carnot engine applied by Ramsey [1]:

$$\eta = 1 - \frac{Q_2}{Q_1} = 1 - \frac{T_2}{T_1} .$$
(12)

Here various results of existence of the machine were discussed in order to be not in contradiction to the principle of increasing entropy.

For an equal-temperature process, there is a simple result:

dS=(dU+PdV)/T,
(13)

where U is the internal energy of system. A general case is (dU+PdV)>0, dS>0 for usual temperature T>0; dS<0 if T<0. Further, if T>0 and (dU+PdV)<0, for example, a contractive process is dV<0, dS<0 is possible [6].

In fact, so long as dS<0, the negative-temperature is unnecessary. Otherwise, "one of the standard formulations of the second law of thermodynamics must be altered to the following: It is impossible to construct an engine that will operate in a closed cycle and prove no effect other than (1) the extraction of heat from a positive-temperature reservoir with the performance of an equivalent amount of work or (2) the rejection of heat into a negative-temperature reservoir with the corresponding work being done on the engine." The experimental study requires that the spin system be well isolated from the lattice system. This isolation is possible if the ratio of spin-lattice to spin-spin relaxation times is large [1].

The condition in which there are more atomic systems in the upper of two energy levels than in the lower, so stimulated emission will predominate over stimulated absorption. This condition may be described as a negative temperature.

In a word, negative temperature is a remarkable question, in particular, for nonequilibrium thermodynamics. Is it a fallacy? From Kelvin scale one obtained infinite temperature and negative temperature, which is inconsistent with other definition of temperature, and with some basic concepts of physics and mathematics. Moreover, "negative temperature" is confused easily with an absolute zero defined usually by negative 273.16C.

## 3. Some Possible Examples for Decrease of Entropy

For a mixture of the ideal gases, the increased entropy is

$$dS = -\sum_{j=1}^{m} n_j R \ln x_j .$$
(14)

If the interactions of two mixed gases cannot be neglected, the change of the free energy will be [9]:

$$G_f - G_i = RT(x_1 \ln x_1 + x_2 \ln x_2) + \lambda x_1 x_2, \tag{15}$$

where $x_1 = n_1 / (n_1 + n_2)$, etc. Then the change of entropy of mixing will be

$$dS = -[\partial(dG)/\partial T] = -R(x_1 \ln x_1 + x_2 \ln x_2) - [\partial(\lambda x_1 x_2)/\partial T]. \tag{16}$$

When $\lambda > 0$, i.e., the interaction is an attractive force, there is probably dS<0. For instance, for $x_1 = x_2 = 1/2$,

$$dS = R\ln 2 - (\partial \lambda / 4 \partial T). \tag{17}$$

When $\lambda > (4R\ln 2)T$, dS<0 is possible [8].

Many protons mix with electrons to form hydrogen atoms, a pair of positive and negative ions forms an atom, and various neutralization reactions between acids and alkalis form different salts. These far-equilibrium nonlinear processes form some new self-organize structures due to electromagnetic interactions. They should be able to test increase or decrease of entropy in isolated systems.

The total rate of production of entropy is [11]:

$$\left(\frac{dS}{dt}\right)_{total} = J_Q \left(\frac{1}{T_1} - \frac{1}{T_2}\right). \tag{18}$$

If the heat current $J_Q > 0$, the total rate will be dS/dt>0 for $T_1 < T_2$, and dS/dt<0 for $T_1 > T_2$. We discussed an attractive process based on a potential energy

$$U^i = -\frac{A}{r}, \tag{19}$$

in which entropy decreases [6].

Using a similar method of the theory of dissipative structure in non-equilibrium thermodynamics, we derived a generalized formula, in which entropy may increase or decrease, the total entropy in an isolated system is [6]:

$$dS = dS^a + dS^i, \tag{20}$$

in which $dS^a$ is an additive part of entropy, and $dS^i$ is an interacting part of entropy. Further, the theory may be developed like the theory of dissipative structure. Barbera discussed the principle of minimal entropy production, whose field equations do not agree with the equations of balance of mass, momentum and energy in two particular cases. The processes considered are: heat conduction in a fluid at rest, and shear flow and heat conduction in an incompressible fluid [13].

Eq.(4) is similar to a generalization of the Boltzmann-Gibbs entropy functional proposed by Tsallis [14], which given by a formula:

$$S_q = -k \sum_i P_i^q \ln_q P_i. \tag{21}$$

where $P_i$ is the probability of the ith microstate, the parameter q is any real number,

$$\ln_q f = (1-q)^{-1}(f^{1-q}-1), (f>0) \tag{22}$$

When q→1, it reduces to

$$S = -k\sum_i P_i \ln P_i . \tag{23}$$

The entropy of the composite system $A \oplus B$ verifies

$$S_q(A \oplus B) = S_q(A) + S_q(B) + (1-q)S_q(A)S_q(B) . \tag{24}$$

Our conclusions are consistent quantitatively with the system theory [6], in which there is [15]

$$S(\rho) \leq S(\rho_1) + S(\rho_2) . \tag{25}$$

This corresponds to Tsallis entropy [14]:

$$S_q = k\frac{1-\sum_{i=1}^{w}P_i^q}{q-1} , \tag{26}$$

which is nonextensive for Eq.(26) when q>1. Both seem to exhibit decrease of entropy with some internal interactions.

For more general case, statistical independence in mathematics corresponds to the independence of probability, i.e., addition of probabilities is $P = \sum_i p_i$. But, addition of dependent probabilities is

$$P(A_1 \cup A_2 \cup .... \cup A_n) = \sum_{i=1}^{n} p(A_i) - \sum p(A_iA_j)$$
$$+ \sum p(A_iA_jA_k) + .... + (-1)^{n-1}\sum p(A_1A_2....A_{n-1}). \tag{27}$$

Therefore, from independence to interrelation, probability decrease necessarily, whose amount is determined by interaction strength. Correspondingly, entropy on mixture of different systems should decrease. In fact, any internal interaction in a system increases already relativity and orderliness.

In present theory, the superfluid helium and its fountain effect must suppose that the helium does not carry entropy, so that the second law of thermodynamics is not violated [9]. It shows that the superfluids possess zero-entropy, but it cannot hold because zero-entropy corresponds to absolute zero according to the third law of thermodynamics. For the liquid or solid $He^3$ the entropy difference [9] is

$\Delta S = S_l - S_s$ >0 (for higher temperature), =0 (for T=0.3K), <0 (for lower temperature).

Such a solid state with higher entropy should be disorder than a liquid state in lower temperature!

Otherwise, in chemical thermodynamics, the entropy of formation is a variant with pressure. In general, any chemical reaction can take place in either direction.

In a word, according to the second law of thermodynamics, all systems in Nature will tend to "heat death" [16], while product will be impossible. But, world is not pessimistic always. The gravitational interactions produce various ordered stable stars and celestial bodies. The electromagnetic interactions produce various crystals and atoms. The stable atoms are determined by electromagnetic interaction and quantum mechanics. According to the second law of thermodynamics, they should be unstable. The strong interactions produce various stable nuclei and particles. A free proton is stable, which testifies that some quarks may form a structure by strong interaction. Theses stabilities depend mainly on various internal interactions and self-organizations. These attractive interactions correspond to decrease of entropy in our theory [6]. Increase of entropy corresponds to repulsive electromagnetic interactions with the same changes and weak interactions. Any stable objects and their formations from particles to stars are accompanied with internal interactions inside these objects, which have implied a possibility of decrease of entropy. The stability in Nature waits our study and research.